\documentstyle[preprint,aps,prl]{revtex}
\begin{document}

\draft
%\twocolumn[\hsize\textwidth\columnwidth\hsize\csname @twocolumnfalse\endcsname

\title{Gel-Electrophoresis and Diffusion of Ring-Shaped DNA }
\author{Uri Alon, David Mukamel}
\address{Department of Physics of Complex Systems, \\
         The Weizmann Institute of Science, 
         Rehovot 76100, Israel
}

\date{preprint}
\maketitle

\begin{abstract}
A model for the motion of ring-shaped DNA in a gel is introduced
and studied by numerical simulations and a mean-field approximation.
The ring motion is mediated by finger-shaped loops (hernias)
that move in an amoeba-like fashion around the gel obstructions.
This constitutes an extension of previous reptation tube treatments. 
It is shown that tension is essential for describing the dynamics in
the presence of hernias.
It is included in the model as long range interactions over 
stretched DNA regions.
The mobility of ring-shaped DNA is found to saturate
much as in the well-studied case of linear DNA. 
Experiments in polymer gels, however,
 show that the mobility drops exponentially with the DNA ring size.
This is commonly attributed to dangling-ends in the gel 
that can impale the ring.
The predictions of the present model are expected to apply  to artificial 2D obstacle arrays
 (W.D. Volkmuth, R.H. Austin, Nature 358,600 (1992)) which have no dangling-ends.
In the zero-field case an exact solution of the model 
steady-state is obtained, and quantities such as the average ring size are calculated.
An approximate treatment of the ring dynamics is given, and the diffusion coefficient
is derived.
The model is also discussed in the context of spontaneous 
symmetry breaking in one dimension.
\end{abstract}

% ]

%\renewcommand{\theparagraph}{\Alph{paragraph}}
%\input epsf

%-------------------- Main Text

{\bf 1. Introduction}

Gel-electrophoresis is a widely used technique for separating DNA fragments according
 to size \cite{mbc}. The separation resolution is limited by a saturation of the mobility at large DNA
 size. Separation of large DNA fragments has been made possible by pulsed-field
 gel-electrophoresis \cite{schwartz,carle}. In view of the phenomenal successes
 of these techniques, an analytic approach to the basic underlying motion of the molecule 
through the gel is desirable.

 Most theoretical treatments \cite{lumpkin,slater,doi,rubinstein,duke prl,duke j chem phys,biopolymers}
 of the motion of the DNA through the gel are based on the reptation 
concept \cite{de Gennes}. 
The DNA is pictured as moving through an impenetrable tube defined by the surrounding gel
 obstructions, with the motion mediated by a snake-like reptation of the polymer ends.  
Reptation
 has proven very successful in describing equilibrium dynamics of polymers in
 gels and melts. Simulations \cite{deutsch} and experiments \cite{gurrier}, however,
 have indicated that for sufficiently long chains undergoing electrophoresis, an
 alternative mechanism of motion is important: the formation of finger-like  hernias 
or leaks through the reptation tube. These hernias (sometimes also called hair-pins, loops or
 kinks) constitute a protrusion of the DNA chain through the walls of the reptation tube 
in a doubled-up loop. Hernias have been included in some recent simulations of linear DNA
 fragments undergoing gel-electrophoresis \cite{zimm,smith,Duke prl 91}.
An additional important effect, that is often neglected in treatments inspired by
equilibrium reptation theory, is tension transmitted along the DNA chain \cite{deutsch}.
 Under a driving 
electric field, strong tension forces
can dramatically affect the polymer motion \cite{Duke prl 91,volkmuth}.

In linear DNA chains, both hernia motion and ordinary  reptation of the chain ends are possible.
In order to separate out and emphasize the effect of hernias, we consider here
 DNA in the shape of a ring (open-circular DNA 
\cite{open circular}).
 The DNA ring can move
 around the gel obstacles only by hernias, sending out fingers in an amoeba-like fashion.
There have been no theoretical studies on gel-electrophoresis
 of open-circular DNA, despite the fact that in practical applications, 
ring shaped DNA (plasmids) is often analyzed by gel-electrophoresis and shows
 behavior different from that of linear DNA fragments \cite{mickel,serwer,levene,wa}.

The behavior of ring polymers in the absence of an electric field is also of interest
 \cite{klein,cates,obukhov}. This problem is related to the behavior of a melt of ring
 polymers, and also to electrophoresis in the weak-field limit through an Einstein 
relation.
 The diffusion of ring-shaped polymers in a lattice of obstructions has been treated by 
numerical simulations and theoretical arguments \cite {obukhov}. An exact treatment of the
 statics and especially the dynamics of ring-shaped polymers in zero field is, however, not
 available.

In this work, a model for the motion of ring-shaped DNA in a gel is introduced
and studied numerically and analytically.
The ring motion is mediated by finger-shaped loops (hernias)
that move in an amoeba-like fashion around the gel obstructions.
This model, described in Sec. 2, constitutes an extension of 
previous reptation tube treatments. 
It is instructive to first study the model neglecting
the effects of tension transmitted along the DNA polymer. Monte-Carlo simulations
of the model, summarized in Sec 3a, show that the chain mobility in this case decreases 
exponentially with DNA size. This is due to the formation of hooks 
which reduce the mobility. This behavior is modified when tension is 
taken into account. In Sec. 3b, tension is added to the model 
as long-range interactions over stretched regions of the chain.
Tension increases the unhooking rates and stabilizes a ring conformation aligned 
with the field. 
This causes the mobility of long ring-shaped DNA to saturate
to a finite value, much as in the well-studied case of linear DNA. 
Experiments in polymer gels \cite{mickel,serwer,levene,wa}, however,
 show that the mobility drops to zero with the DNA ring size, with large
rings hardly penetrating into the gel. This
is commonly attributed to the rings becoming impaled on dangling-ends in the gel.
The predictions of the present model are expected to apply 
to artificial 2D obstacle arrays
 \cite {austin} which have no dangling-ends. In Sec 3c, the polymer motion 
is qualitatively described by a mean-field treatment.
In the zero-field case, discussed in Sec. 4, an exact solution of the model 
steady-state is obtained, and quantities such as the average ring size
are calculated.  In Sec 4b, an approximate treatment of the
zero-field ring dynamics is given, and the diffusion coefficient is derived.
 This gives an analytic foundation to previous scaling arguments \cite {obukhov},
 and suggests a framework
for analysis of dynamical features of driven polymers.
In Sec. 5, the model is also discussed in the context of spontaneous
 symmetry breaking in one dimension.

\noindent
{\bf 2. Model for DNA in a Gel Including Hernias}

We present a model for a charged polymer ring reptating in an electric field in an array
 of obstacles (eg. a gel). The model is  based on the Rubinstein-Duke (RD) approach 
\cite{rubinstein,duke prl,duke j chem phys}, and is extended here to 
include hernias, which are hair-pin shaped excursions out of the usual reptation tube.
 Hernias are crucial for polymers in the shape of a ring, 
in which the motion through the surrounding obstacles may be accomplished only by
 hernia fingering.

We begin by describing the RD model for reptating linear polymers. We then extend the model to 
include hernias. In the RD model,
the gel is idealized as a lattice of point obstacles, with pore diameter $b$, as shown
 in Fig 1. In agarose gels, $b \sim 100nm$, while in recently introduced artificial
 obstacle arrays $b \sim 1 \mu m$ \cite{austin}. The DNA is represented as a chain
 of $L$ segments, each of one persistence length ($\sim 50nm$). The segments may either
 be stretched, when the polymer threads through adjacent pores, or coiled, when
 the segment is contained in one pore.  Each configuration of the polymer is
 represented by the positions of the successive cells that the polymer threads.
A simplified description of the chain is
 coded by the projection along the field direction of the
 displacement between segment ends. This displacement can be either $+b$, when the
 segment threads between two pores in the field direction, $-b$ when it threads
 two pores against the field direction, or $0$ when the segment is coiled in one
 pore. Thus the configuration is reduced to a 1D
 lattice of $L$ sites. Each site $i$ corresponds to a DNA segment, and
 has a state $\phi_i$, which can be either $+$, $-$ or $0$, as demonstrated in Fig 1.
Note that in this description some information regarding the microscopic configuration is lost.
However, it provides a convenient way to model the dynamics 
\cite{rubinstein,duke prl,duke j chem phys}.

The chain reptates by the motion of the coiled, lax segments through the chain.
 In aquaeos solution, the DNA is assumed to be  uniformly charged. The forces acting
 on each segment are an electric force $F_e=QE$ where Q is the charge per segment and
 $E$ is the field strength, and a thermal Brownian noise $F_{th}$ of the order of
 $kT/b$. These forces are represented in the model by the following rules. At each
time step, a pair of sites is chosen at random, and a move is made with the following rates:
\begin{eqnarray}
+0 \rightarrow 0+, \; \mbox {at rate q,} \; \; 0+ \rightarrow +0 \; \mbox {at rate p} \label{eq p}\\
-0 \rightarrow 0- \; \mbox {at rate p,} \; \; 0- \rightarrow -0 \; \mbox {at rate q} \label{eq q}
\end{eqnarray}
while $+-$ or $-+$ pairs are stuck, since they represent two stretched segments hooked
 around a gel obstacle (see Fig. 1). Moves in the field 
direction are favorably biased over the reverse moves, through the rates $p$ and $q$.
 These rates are determined, in the case of weak fields, by local detailed-balance 
conditions \cite{duke prl}, such as
\begin{eqnarray}
p=\omega_0 \exp(\epsilon/2) \\ 
q=\omega_0 exp(-\epsilon/2)
\end{eqnarray}
where $\omega_0$ is a microscopic rate and $\epsilon=QEb/kT$.
Note that this is a nonequilibrium dynamics and it does not
obey full detailed-balance. 
The ratio between these rates is thus a Boltzmann factor of the ratio of
 electrical to thermal energy, 
\begin{equation}
p/q=exp(\epsilon) \;\;\;\; \epsilon=QEb/kT
\label{eq detailed balance}
\end{equation}
These rules, along with rates for injection of $+$ and $-$ at the head and tail of
 the linear chain, define the Rubinstein-Duke model 
\cite{duke prl,duke j chem phys}.

We now extend this model to account for hernias. A hernia amounts to a projection 
of the chain from a pore with at least 2 coiled segments (2 adjacent $0$ sites) 
into a new pore, threading
 one segment out of the pore and another segment back to the original pore (for
 example move B in Fig. 1). This corresponds to a pair creation move $00 \rightarrow +-$.
The reverse annihilation move $+- \rightarrow 00$ corresponds to the hernia retracting
and forming two coiled segments in a single pore. After a pair is created,
the $+$ and $-$ can diffuse according to the RD rules. An important feature of the model
is that pairs of $+$ and $-$ which are created together are tracked as a connected
pair throughout the dynamics. Each $+$ in the configuration has a unique $-$ to 
which it is connected.
 Keeping track of such connections between pairs is necessary in order 
to track the hernia finger hierarchy.
To see this,
consider a pore with many coiled segments. A number of hernias may be formed, 
projecting into different neighboring pores.
 An important point is that $+$'s and $-$'s from different hernias can not
 annihilate (assuming that several hernias originating from the a pore always 
project to different neighbors, a reasonable assumption for pore lattices of
 high coordination number). Thus, the pairings of $+$'s and $-$'s must be tracked:
 each $+$ can annihilate only with the unique $-$ to which it is paired.  
Each configuration is defined by the $+$ and $-$ and $0$ sites, along with 
their pairing to hernias (Fig. 2). Only pairings in which the
hernia pairs are nested are allowed, as shown in Fig. 3 (pairing lines may not cross each other).
 Starting with an allowed configuration, the hernia creation rules assure that the configuration
at each subsequent time is also allowed. The phase space is larger than 
in the original RD model which has the three states $+$,$-$ and $0$ for each site,
 but no pairings.

The hernia creation and annihilation moves, which supplement the reptation moves of Eqs.
\ref {eq p} and \ref {eq q}, are: 
\begin{eqnarray}
00 \rightarrow +_{\sqcup}-, \; \mbox {at rate c,} \; 
\; 00 \rightarrow -_{\sqcup}+ \; \mbox {at rate $c'$} \\
+_{\sqcup}- \rightarrow 00, \; \mbox {at rate a,} \;
 \; -_{\sqcup}+ \rightarrow 00 \; \mbox {at rate $a'$} 
\end{eqnarray}
The symbol $_\sqcup$ denotes pairs of $+$ and $-$ that have been created together 
(connected pair). 
The hernias tend to grow, as the field bias pushes the $+$'s to the left and $-$'s 
to the right.
 The hernias may develop sub-hernias, and a hierarchy of hernias may form. 
An example is shown in Fig. 2, for a ring shaped polymer.
Thus, the polymer can assume a highly ramified shape, with a hierarchy of hernias of 
different sizes.

The model as described above neglects an important physical effects: 
the tension transmitted along the chain.
 As shown below, this proves to be very important in the electrophoresis of ring shaped DNA.
 Tension acts as a long range effective interaction, and is included in the model as
 described in Sec 3b.

\noindent
{\bf 3 Gel Electrophoresis of Open-Circular DNA}

Using the model, we studied gel-electrophoresis of ring-shaped 
(open circular \cite{open circular}) DNA. Periodic boundary conditions are thus imposed in
 the model. The ring is not concatenated with any gel obstacle (it is prepared
 outside the gel and moves into the gel under the field influence). We first study the
model in the absence of tension by
  Monte-Carlo simulations.  The treatment of tension, and its
effects on the dynamics are presented in Sec 3b. A simple mean-field treatment
is given in Sec 3c.
\\
{\bf 3a. Monte-Carlo results:}
It is instructive to first study the model as described in Sec 2, neglecting
the effects of tension. In order investigate the model, we performed Monte-Carlo simulations,
 typically using
$a=c=1$, $a'=p$,$c'=0$, $p/q=1.01-2$, and ring lengths up to $L=100$. The mobility as
 a function of time is shown 
in Fig. 4. It is seen, that the mobility
 displays a spiked behavior: the system is effectively
 in one of two states, one with a positive mobility, and one with a zero mean mobility. 
The average lifetime of the
 zero-mobility states grows as $L$ increases.

The nature of the dynamics is clarified by snapshots of the ring configurations in the
 two states, shown in Fig. 5. It is seen that the ring cycles between 
quasi-linear and hooked states. The high-mobility phases corresponds to the quasi-linear 
conformation in which the ring is aligned with the field \cite{quasi}. The conformation of
effective charges in the model that represents this conformation is shown in Fig. 6a.
This conformation is short lived, 
because it develops an instability: a hernia that buds on the side of the quasi-linear
 ring grows into a hooked configuration with two stretched arms, pinned over an obstacle (see Fig. 6b).
 In this phase, the ring is stuck, and there is zero mean mobility. The hooked state persists 
for a long time. The ring unhooks by one of the arms retracting by fluctuations,
 until a new, quasi-linear, high mobility shape is attained. 
This explains the burst-like structure of the mobility.  In the absence of tension, 
the mean mobility $\mu$ decreases exponentially with $L$ (Fig. 7). This
 is because the unhooking rate, by which an unstretched segment moves from one arm to 
the other, is exponentially small, since it takes of order $L$ steps against the field 
for the segment to escape the arm's effective potential trap. Similar dynamics
 can occur also in linear chains \cite{deutsch,Duke prl 91}. In the unhooking phase, 
tension plays a crucial role, as described in the following section.

{\bf 3b. Effect of Chain Tension:} It is important to consider the effects of
 tension transmitted along the chain \cite{deutsch,Duke prl 91,volkmuth,sevick}.
 The main effect of tension is to dramatically increase the unhooking rates of
 stretched hooks. It acts as a long-range interaction between coiled segments
 \cite{Duke prl 91}. Tension in the context of gel-electrophoresis of linear DNA
was treated in a previous study \cite {Duke prl 91}, where coiled segments 
were allowed to make long-ranged hops along the chain. The present treatment
 simplifies this by using only local hops. In addition, the present model extends
Ref \cite {Duke prl 91} by taking into account the effect of tension on the hernia 
creation and annihilation rates. 

In order to model the effect of tension, we note
that under the influence of an electric field the charged chain behaves
like a chain moving in a gravitational field coupled to its weight.
The tension transmitted along the chain is relaxed at coiled (0) segments and at hernia tips 
($+_\sqcup-$ and $-_\sqcup+$ paired at neighboring sites). Note that unpaired
neighboring $+-$ or $-+$ sites represent segments of DNA chain which are draped over
a gel obstruction and therefore are capable of transmitting tension.
We define an effective field for each pair of sites, which corresponds to the tension generated 
by regions of stretched chain on both sides of the sites. This results in a 
movement rate $\omega_j$
for the pair of sites $j$ and $j+1$, which depends 
on long-range interaction between different sites, as described below. 
At each step of the simulation, a pair of sites $j$ and $j+1$ is chosen at random.
Pairs at which there is an allowed move are of three types: (a) coiled site adjacent to
a stretched site $\phi_j,\phi_{j+1}=0-,-0,0+$ or $+0$, (b)
 two coiled sites $\phi_j,\phi_{j+1}=00$ (creation move) or (c) a hernia tip $\phi_j,\phi_{j+1}=+_\sqcup -$ or
$-_\sqcup +$ (annihilation move). We will refer to such pairs as {\it relaxed pairs}.
If the pair is not one of these three types, it remains unchanged.
If the pair has an allowed move, the move is performed with the rate $w_j$, and the states
of sites $j$,$j+1$ are accordingly adjusted. A new 
pair is chosen and the process is repeated.

To derive the movement rate for pair $j$ (site $j$ and $j+1$), 
$\omega_j$, we consider the tension transmitted along the
DNA due to stretched regions of chain on either side of the pair.
Since the tension accumulates
 along these regions, the local tension field is proportional to the net displacement in the field 
direction of these stretched regions. The stretched regions terminate at either a
 coiled segment ($0$ site) or a hernia tip since theses are points when the chain tension
 is relaxed. 
The effective tension force acting on a pair consisting of a coiled segment adjacent to a stretched one
 $(0,\phi_{j+1})$, with $\phi_{j+1}=\pm 1 $, is
\begin{equation}
F_j=\frac12 \epsilon \sum_{m=j+1}^{k_1} \phi_m \label {eq F1}
\end{equation}
where $k_1$ is the closest succeding site to site $j+1$ which is a member of a relaxed pair,
and the dimensionless external field is $\epsilon=QEb/kT$.
Similarly, for a ($\phi_j$,0) pair with $\phi_j=\pm 1$, the effective force is given by
\begin{equation}
F_j=-\frac12 \epsilon \sum_{m=k_2}^{j} \phi_m \label {eq F2}
\end{equation}
where $k_2$ is the closest preceding site to site $j$ which is a member of a relaxed pair.
The force acting on a hernia tip ($+_\sqcup-$ or $-_\sqcup+$ pairs) is
\begin{equation}
F_j=\frac12 \epsilon (\sum_{m=j+1}^{k_1} \phi_m-\sum_{m=k_2}^{j} \phi_m) \label {eq field}
\end{equation}
with similar definitions of $k_1$ and $k_2$.
(We note that the model can also be applied to linear chains,
where additional points at which tension is relaxed are the chain ends).

As an example, consider the  configuration of Fig 2. This configuration contains one coiled segment 
at site $i=10$, $\phi_{10}=0$. Consider sites 9 and 10, at 
which the configuration is $-0$. To evaluate 
the effective field for this pair, we sum $\phi$ in the two sites to the right of the pair 
(these sites cancel each other),
where we reach a hernia tip. The total
effective force is $F_9=0$. At sites $10$ and $11$, where the configuration is $0+$,
tension accumulates along a three sites stretched $+$ region to the right of the pair, which
terminates at a hernia tip, and $F_{10}=\frac32 \epsilon$. 
 
  The movement rate $\omega_j$ can be related to the local tension force $F_j$
from a consideration of the thermal
 and friction forces on the string. The motion of the DNA segments through
 the solvent is such that viscous drag forces are much larger than any inertial term
 \cite {deutsch,volkmuth}.
The behavior of the chain under thermal noise can be treated using a 
Fokker-Planck approach \cite {Duke prl 91}, using a Smoluchowski equation for
 the string motion along the tube contour, which includes a Brownian term and a
 friction coefficient proportional to the string's length. In the present work we 
propose a simpler physical model, which is valid at both the strong and weak field
 limits:
\begin{equation}
\omega_j=\cases{r_j\exp(F_j) & $F_j<0$ \cr
                r_j(1+F_j) &   $F_j>0$}
 \label {eq string}
\end{equation}
where $r_j$ is equal to a microscopic rate $\omega_0$ for pairs where a coiled segment can
 move ($\phi_j, \phi_{j+1}=+0,0+,0-,-0$), $r_j=c_0$ for pairs where a hernia
 can be created ( $\phi_j, \phi_{j+1}=00$) and  $r_j=a_0$ where a hernia
 can be annihilated  ($\phi_j, \phi_{j+1}=+_\sqcup-,-_\sqcup+$ ). At all other
 pairs of sites, $r_j=0$, since no other moves are allowed. The constants
 $\omega_0$,$c_0$ and $a_0$ are the microscopic rates of the various processes.
 Eq. \ref {eq string} goes to 
a Boltzmann factor for low effective fields where it represents local detailed-balance. 
At high effective field strength, the movement rate becomes linear in the effective field strength.
 This is expected since at large fields thermal fluctuations become unimportant
 and the local chain velocity becomes proportional to the local force \cite{deutsch}. 
Eq. \ref {eq string} allows hernia annihilation rates 
to be affected by chain tension, with annihilations at hernia tips flanked by long regions 
stretched in the field direction given a high rate.

The model without tension, described in Sec. 2, can be recovered from this model, by taking 
$k_1=j$ , $k_2=j+1$ in Eq. \ref {eq F1}-\ref {eq field}. This corresponds to the screening
 limit, when the density of coiled segments is so high that there appear no
 extended regions of stretched segments in which tension can develop, and the 
field at each bond is due to the external field alone. In this case, the
 movement rates are related to those of the model without tension described in
 Sec 2 at low fields ($\epsilon<<1$) via: $p=\omega_0 e^{\epsilon/2}$,
 $q=\omega_0 e^{-\epsilon/2}$, $c=c_0$,
 $a=a_0 e^{-\epsilon}$,$c'=c_0$,$a'=a_0e^{\epsilon}$. It is seen
that in this limit, the rates satisfy Eq. \ref {eq detailed balance}.

Tension causes hooks to have a much smaller effect on the mobility. Monte-Carlo
 calculation including tension at two different field strengths are 
shown in Fig. 7. 
 We find that the ring mobility saturates at large DNA sizes, much as in the well
 studied case of linear DNA fragments. It is interesting to note that the ring
 arranges itself into a quasi-linear shape in the size regime studied. Here,
 coiled segments are very frequent - roughly 1/3 of the segments are coiled. Thus,
 {\em tension is screened by the coiled segments} 
and has a very small effect during most of the dynamics, since it is important only in long,
 continually stretched pieces of the chain. Only when a side-hernia forms, and a hook
 begins to be created, does tension come into play, and essentially stabilizes the
 quasi-linear shape aligned with the field. We note that for very large rings, branching
 effects similar to those discovered in large linear fragments in Ref
 \cite {Duke prl 91}, are likely to occur, though the ring mobility should remain
 constant.

The screening of tension by coiled segments is very important in explaining the sucsses of 
 reptation tube models for linear DNA 
which seems to describe experiments on linear DNA fragments quite well \cite {biopolymers,schutz}, 
though the models neglect both hernias and tension.  
The present results suggest that when including hernias in a model of polymer dynamics, it is 
essential to also take
 tension into account, as these two effects have a canceling behavior, respectively
increasing and decreasing the hooking rates. 

The predictions of the present model that the mobility saturates with the ring size
seem to disagree with experiments. Studies of open-circular DNA (plasmids) run through agarose gels
 show that above a certain DNA size, the
 plasmids are "stuck at the wells" and do not enter the gel \cite{mickel,serwer,levene,wa}.
 The explanation offered by Micel et. al. \cite{mickel} is that the rings become hooked on dangling
 ends in the gel (unconnected ends of the gel fibers or other impurities that penetrate the pores),
 which impale the ring ("hoop in stick" effect). The ring can unhook
 by a fluctuation which can overcome the field pulling the ring. The probability of
 such a fluctuation is exponentially small in the ratio between the electric
force pulling the ring and the thermal forces, and the mobility is
\begin{equation}
\mu \sim \exp(-QENb/kT)
\end{equation}
 The saturation of the mobility predicted in the present model could be checked 
experimentally
 on recently introduced artificial 2D arrays of obstacles \cite{austin} with no
 dangling-ends that can impale the ring, as suggested in Sec. 6.

{\bf 3c. Mean-field treatment:}
%say tension is screened - so neglected and simple MF

In the presence of tension, the DNA is found mostly in a quasi-linear conformation
aligned with the field, with
many coiled segments. The coiled segments essentially screen
tension. This allows for a
simple and local mean-field treatment of the DNA motion.  
 
Consider a quasi-linear chain, (Fig. 5, rightmost and leftmost configurations). 
 In this configuration, hernias are annihilated at the upfield end of the ring.
The coiled segments ($0$'s) which are generated move down to the leading end, 
where a new hernia is formed.  The
density of coiled segments, $\rho$, is given by a balance 
of hernia creation and annihilation. The rates for these processes 
are obtained in the mean field approximation by neglecting
 correlations:  Hernias are created (at the leading end) when two coiled segments are adjacent, 
at a rate $c \rho^2$, and
annihilated (at the upfield end) when two stretched segments are adjacent at rate $a(1-\rho)^2$.
 In this approximation for the annihilation rate, 
the assumption that the ring is quasi-linear is used, since a stretched
segment can annihilate only with its unique pair. In a random configuration
the pair would have a small chance of being adjacent. Here we assume that in the quasi-linear 
configuration, a pair of stretched segments at the leading end may always be annihilated.

The balance between annihilation and creation yields
\begin{equation}
\rho=1/(1+\sqrt{c/a})
\end{equation}
The mobility $\mu$, equal to the mean center of mass displacement per unit time,
 is given by an average over all the configurations allowing movement,
 weighted by the respective rates. Since in steady state the hernia 
annihilation and creation moves balance each other, we have, in the simplest 
mean field approximation that the mobility is proportional to the probability of finding
 a stretched segment adjacent to a coiled one:
\begin{equation}
\mu=(p-q) \rho (1-\rho) \; \; \;\;\; \rho=1/(1+\sqrt{c/a}) \label {eq mf mu}
\end{equation}

The  qualitative features of the mobility are reasonably described by the simple 
mean-field theory as shown in Fig. 8, where the
 density of coiled segments and the mobility as a function of the ratio of hernia creation
and annihilation rates $c/a$ are shown. At high ratio of creation to 
annihilation rates $c/a$, the 
chain is dense with stretched segments and the mobility is low. At low $c/a$, there are few
stretched segments that can move, and the mobility is again low. Around $c/a=1$, where
the density of coiled segments is around $1/2$, the mobility is at a peak. The simulations show 
similar qualitative behavior. The mean-field mobility overestimates the full model
mobility by a factor of about 2. This is probably due to processes that impede the motion, 
such as hooking and pair creation in the bulk of the chain and
not only at the head and tail, that are neglected in the mean-field treatment.

\noindent
{\bf 4. The Zero Field Case}

We now turn to the case of zero electric field. This case is important as a 
question in classical polymer physics \cite{klein,cates}: what is the effect 
of hernias on the statics and dynamics of a chain in a gel or melt at equilibrium? 
In addition, 
the zero-field diffusion can be related to the low-field electrophoretic mobility via
 the Einstein-Nernst relations.

The zero-field case offers a significant simplification in the model: tension can be
 ignored in this case, and the model described in Sec 2 is used,
 with $q=p$, $c=c'$ and $a=a'$.
The probability of a given configuration $C$, $P(C)$, is governed by the 
Master equation
\begin{equation}
dP(C)/dt=\sum_{C'} \lbrace W(C' \rightarrow C)P(C')-W(C \rightarrow C')P(C) \rbrace
\end{equation}
where $W(A \rightarrow B)$ is the rate of transition from configuration $A$ to $B$.
 A solution to
 the master equation is found which satisfies detailed-balance. Each move which 
preserves the number of stretched segments, such as $+0 \rightarrow 0+$, 
is exactly balanced by its reverse move. Moves where hernias are created or
annihilated are balanced by the reverse move, with an extra factor related to the 
creation and annihilation rates. A configuration $A$ which has a $00$ at a certain bond,
can, in a single move, go to or be reached from only two configurations $B,B'$ which are exactly
the same as $A$ except that they have either a $+_\sqcup -$ or a $-_\sqcup +$ at the bond.
 The solution for the
 probability of configuration $C$ is
\begin{equation}
P(C)=N(L)^{-1} (c/a)^{h(C)}
\end{equation}
where $c/a$ is the ratio between hernia creation and annihilation rates,
 and $h(C)$ is the total number of hernias in the configuration $C$. 
In the steady state, all configurations have equal probability, up to a factor
 depending only on the total number of hernias in the configuration. 
This solution is remarkable in that although there are strong interactions between
 different hernias, the probability of each configuration depends only on the number
 of hernias and not their relative positions and sizes.
The normalization factor
 $N(L)=\sum_C (c/a)^{h(C)} $ is connected to the number of allowed configurations 
(only configurations with nested hernias are allowed, as shown in Fig. 3). $N(L)$ satisfies the 
recursion relation:
\begin{equation}
N(L)=N(L-1)+(2c/a) \sum_{l=0}^{L-2} N(l)N(L-2-l) \label {eq N rec}
\end{equation}
The terms on the right-hand side can be understood as follows: given a ring of size
 $L$, choose a site. The first term corresponds to the case where the site contains
 a $0$, and thus the configurations of hernias can be mapped to a ring of size 
$L-1$ by deleting the site.
The second term is the case where the site is a member of a hernia with its pair
 at a distance of $l+1$ sites (hernia of size $l$). The factor $2$ in Eq. \ref {eq N rec}
 is due to the two possible assignments of $+$ and $-$ charges to a hernia pair, which
have equal probability in the absence of an electric field. 
The values $N(0)=N(1)=1$ are supplemented to this recursion relation. 
 Solving for the asymptotic
 form for $N(L)$ at $L \gg 1$, we consider the function $u(L)=\eta^{-L} N(L)$.
For sufficiently large $\eta$ it has a finite integral.  Using a Laplace transform of
$u(L)$, $g(s)=\sum_{L=0}^{\infty}u(L)e^{-sL}$, in the recursion relation
 Eq. \ref {eq N rec}, we find \cite {sign}
\begin{equation}
g(s)=\frac{1-e^{-s}\eta^{-1} +\sqrt{(e^{-s}\eta^{-1}-1)^2-4(2c/a)e^{-2s}\eta^{-2}}}
{2(2c/a)\eta^{-2} e^{-2s}}
\end{equation}
The smallest value of $\eta$ for which $g(0)=\sum u(L)$ exists is 
$\eta=2\sqrt{2c/a}+1$.
At this value of $\eta$, for small $s$, 
$g(s) \approx g_0+g_1s^{1/2}$. This
corresponds to the following asymptotic form of the partition sum at $L \gg 1$
\begin{equation}
N(L) =N_0 L^{-3/2} (2 \sqrt{2c/a}+1)^ L \label{eq Nasy}
\end{equation}
with 
\begin{equation}
%N_0 =(4 \sqrt{\pi}c/a)^{-1}(1+2\sqrt{2c/a}) [\sqrt{2c/a}+4c/a]^{1/2}
N_0 =(4 \sqrt{\pi}c/a)^{-1}(1+2\sqrt{2c/a}) [\sqrt{2c/a}+4c/a]^{1/2}
\end{equation}
This allows derivation of steady state densities, such as $\rho$, the density of
 coiled segments (0's). This density
is given from the construction of
 the recursion relation
Eq \ref{eq N rec} simply by those configurations at which
a given site is $0$, compared to the total weight of the configurations:
\begin{equation}
 \rho=N(L-1)/N(L)=(2 \sqrt{2c/a}+1)^{-1}
\end{equation}
The hernia size distribution, $n(l)$, defined as the probability that a selected
 site belongs to a hernia of size $l$, is $n(l)=N(l)N(L-2-l)/N(L)$ (see Eq \ref{eq N rec}). Thus, for $1 \ll l \ll L/2$,
\begin{equation}
 n(l) \approx N_0 l^{-3/2}
\end{equation}
This suggests that the ring polymer adopts a ramified fingered shape,
 with a power-law spectrum of finger sizes.

{\bf 4a. Average Ring Size:} The exact solution also allows calculation of the mean size 
of the ring at zero field.
 The radius of gyration serves as a convenient measure of the ring size: 
it is defined as the mean squared distance between two sites 
(i.e. the sum of number of $+$ sites minus the
 number of $-$ sites between $i$ and $j$ squared, averaged over all
 configurations 
and all $i$ and $j$).  To evaluate the radius, we define the function 
\begin{equation}
G(i)=\sum _C R^2_C(i) P(C)
\end{equation}
where 
\begin{equation}
R^2_C(i)=(\sum_{k=1}^{i}\phi_k)^2
\end{equation}
where $\phi_k$ is the state of site $k$ in configuration $C$.
 That is, $G(i)$ is the average over all
 configurations of the squared distance between the two "test sites" 
$1$ and $i$. 
The radius of gyration $R$ is simply given by the root of the mean of $G(i)$:
\begin{equation}
R^2=L^{-1}\sum _{i=1}^L G(i)  \label {eq R def}
\end{equation}
We construct a recursion relation for $G(i)$, noting that the only contribution
that does not average to zero is from unclosed hernias in the interval
 between the sites. Since there is no field, each site on a hernia can be
a $+$ or a $-$ with equal probability. The mean squared distance, averaged over all assignments of 
$+$ and $-$ to the hernias, is just the number of unclosed hernias. 
To build a recursion relation for $G(i)$, we go 
from an interval of size $i-1$ to size $i$. Thus, to the interval of sites $1...i-1$
we add one site, $i$, which may 
be either a $0$, belong to a hernia that closes outside the interval,
 or close one of the open hernias in the interval. Thus, we divide all of
the possible configurations into three groups: $C_0$ in which site $i$ is in a 0 state, $C_1$ in
which site $i$ belongs to a hernia that closes outside the interval, and
configurations $C_2$ in which site $i$ pairs with a site inside the interval, 
thus closing a hernia. In configurations $C_0$, $R_C^2(i)=R_C^2(i-1)$. In
configurations $C_1$, $R_C^2(i)=R_C^2(i-1)+1$. In
configurations $C_2$, $R_C^2(i)=R_C^2(i-1)-1$. Thus,
\begin{equation}
G(i)=\sum _C R^2_C(i-1) P(C)+\sum_{C_1}P(C)-\sum_{C_2}P(C)
\end{equation}
This leads to the recursion relation
\begin{eqnarray}
G(i)=G(i-1)+(2c/a)[\sum_{l=i-1}^{L-2}N(l) N(L-2-l) \label{eq G rec} \\
-\sum_{l=0}^{i-2}N(l) N(L-2-l)]/N(L) \nonumber
\end{eqnarray}
The boundary condition is $G(1)=1-N(L-1)/N(L)$, since the mean squared displacement due to a single site
is just the density of uncoiled segments $1-\rho$. $G(i)$ rises to a peak when $i=L/2$ 
and then drops off to $0$ at $i=L$, because the ring is closed and going around the ring the mean
displacement must go to zero. Going to the
 continuum limit, and using the symmetry of $G(i)$ around
$i=L/2$, we find
\begin{equation}
dG(i)/di=(2c/a) N(L)^{-1}\int_i^{L-i} N(l)N(L-l)dl
\end{equation}
For large $i$, using the asymptotic form of $N(L)$ found above (Eq. \ref {eq Nasy}),
 the integrals involved in calculating $G(i)$ can be performed analytically.
 This yields
\begin{equation}
G(i)=8 (2c/a) N_0 \sqrt {i}\sqrt {L-i} / \sqrt{L}
\end{equation}
 showing that the mean maximal excursion of the
 ring is $G(L/2) \propto \sqrt{L}$. Integrating over $G(i)$, one obtains
\begin{equation}
R =\sqrt{\pi (2c/a) N_0} L^{1/4}
\end{equation}
This scaling is valid for long chains.
 The radius for finite chains can 
be readily found from
 the recursion relations \ref {eq N rec} and \ref {eq G rec}.
 Thus, the ring adopts a much more compact configuration than the linear, 
reptating chain, in which $R \propto L^{1/2}$. This form is valid for chains small enough that
 excluded volume effects can be neglected \cite {obukhov,parisi}. 
An intuitive argument predicted
 the $R \sim L^{1/4}$ scaling \cite {stockmayer,kholkov}, 
by mapping the ring to a randomly branched
 graph. The exact solution of present model allowed us to derive this 
scaling exactly, from a steady state of a dynamic model.

{\bf 4b. Dynamics of a Ring: } The solution for the steady state allows a rather accurate approximation for
 the polymer dynamics. We consider the motion of a marked hernia on the ring. We define
 $P(l,t)$ as the probability of finding the hernia at size $l$ at time $t$. The hernia 
can grow if there is a $0$ adjacent to the hernia from the outside, and it can shrink if 
there is a zero adjacent to the hernia from inside. We make the approximation that these 
probabilities are given by the corresponding steady-state probabilities. This
 approximation is a good one for large hernias. This is because the motion of the
 hernias is diffusive, and hence many configurations of the ring segments inside and 
outside the marked hernia are sampled on the time-scale of the effective hernia motion. 
The evolution equation for the size of the marked hernia $P(l,t)$ is
\begin{eqnarray}
\partial P(l,t)/\partial t=\rho_i(l+1) P(l+1,t)+\rho_o(l-1)P(l-1,t)\\
-[\rho_i(l)+\rho_o(l)]P(l,t) \label {eq plt1} \nonumber
\end{eqnarray}
where $\rho_i(l)$, $\rho_o(l)$ are the probability of a zero adjacent to the size-l
 hernia from the inside and outside respectively (a microscopic rate constant has been
 factored out of the equation so that each move take one time unit).
At the boundaries, the equations have terms corresponding to the annihilation of the hernia, 
which is possible when it is of size $l=0$ or $l=L-2$:
\begin{eqnarray}
\partial P(0,t)/\partial t = \rho_i(1) P(1,t)-[a+\rho_o(0)]P(0,t) \label {eq bnd}\\
\partial P(L-2,t)/\partial t = \rho_o(L-3) P(L-3,t)\\  \nonumber
-[a+\rho_i(L-2)]P(L-2,t) \nonumber
\end{eqnarray}

The main idea of the present approximation is to use the exact steady state solution to
estimate the probabilities of motion for the hernia.  This yields
\begin{eqnarray}
\rho_i(l) &=& N(l-1)/N(l) \sim \rho (1+\frac32l^{-1} ) \\
\rho_o(l) &=& N(L-l-1)/N(L-l) \sim \rho [1+\frac32(L-l)^{-1} ] 
\end{eqnarray}
where the asymptotic forms are valid at $l \gg 1$ and $L-l \gg 1$, and
 $\rho=(2 \sqrt{2c/a}+1)^{-1}$. Note that smaller hernias have on 
average more zeros inside them than larger hernias. This is an entropic
 effect, due to the larger number of hernia-pairings in a large hernia.
 This creates a bias for small hernias to shrink, an effect that has important
 consequences for the ring dynamics, as shown below.

In order to analyze the scaling of Eq. \ref {eq plt1}, it is convenient to turn to a continuum
 form:
\begin{equation}
\partial P(l,t)/\partial t=\rho \nabla ^{2} P(l,t)+ \rho \nabla [U(l) P(l,t)] \label {eq plt}
\end{equation}
supplemented with appropriate boundary conditions (the exact
form of the boundary conditions do not affect the scaling results obtained below). 
The potential 
\begin{equation}
U(l)=\frac{3}{2}  [l^{-1}-(L-l)^{-1}] 
\label {eq pot}
\end{equation}
corresponds to the bias of small hernias to shrink. Consider the case of a marked
 hernia created at time $t=0$. Thus, the initial conditions are $P(l,t=0)=\delta(l=0)$. 
We define
\begin{eqnarray}
F(l) &=& \int_{0}^{\infty} P(l,t) dt\\
T(l) &=& \int_{0}^{\infty} t P(l,t) dt
\end{eqnarray}
Thus the mean lifetime of the marked hernia is
\begin{equation}
 \tau=\int_{0}^{\infty} T(l) dl/\int_{0}^{\infty} F(l) dl \label{eq tau}
\end{equation}
Ordinary differential equations for $T(l)$ and $F(l)$ may be easily formed by
 appropriately integrating over Eq \ref {eq plt}: 
\begin{eqnarray}
-P(l,t=0) &=& \nabla^2 F(l)+\nabla[U(l)F(l)] \label{eq F}\\
-F(l) &=& \nabla^2 T(l)+\nabla[U(l)T(l)] \label{eq T}
\end{eqnarray}
These equations are exactly soluble, yielding rather complicated expressions. However, to understand
 the scaling form of the solutions, it is useful to consider a simpler potential of the form $U=u_0/l$, 
which is equivalent to the full
potential Eq. \ref {eq pot} at $l<<L$, with $u_0=3/2$. Plugging in power law forms for
 $F$ and $T$ in Eq. \ref {eq F},\ref {eq T},
we find $F(l) \sim l^{-u_0}$ and $T \sim l^{2-u_0}$. For $u_0=3/2$, $\int_0^{\infty} F(l) dl \sim F(0)$ and
$\int_0^{\infty} T(l) dl \sim L^{3/2}$, yielding for long chains
\begin{equation}
\tau \sim \tau_0 L^{3/2} \label{eq tau}
\end{equation}
The same asymptotic result is found using the full potential of Eq. \ref {eq pot}.
The lifetime of a given marked hernia is much shorter than that expected from a
 simple diffusion argument, $\tau \sim L^2$ (which corresponds to using no potential, $U=0$,
 in the above calculation).
 This is due to the entropic bias of small hernias to shrink, which compels hernias to spend less time in 
"random walk" motion as compared to the pure diffusion case. Note that the power law exponent $3/2$
in Eq. \ref{eq tau} is related directly to the prefactor of the effective potential that is derived from the detailed 
steady state solution.

From this result, one can readily derive the scaling of the ring center 
of mass diffusion coefficient, $D$, using a classical scaling argument \cite{de Gennes}. Essentially, diffusion
proceeds by the transport of hernias along the ring. Each hernia takes a time $\tau \sim L^\theta$ to
travel a distance of order $R(L) \sim L^\nu$, the linear size of the polymer. If there
were only one hernia on the chain at any time, the center of mass diffusion constant $D_0$ would
obey 
\begin{equation}
D_0 \tau=(R(L)/L)^2
\end{equation} 
where the right hand side represents the mean square 
displacement of the center of mass arising from the transport of one hernia across a distance $R$.
In reality, the number of hernias present at any time is proportional to $L$.
 Hence the center of mass diffusion constant $D$ scales as 
\begin{equation}
D \sim L D_0 \sim L^{2\nu -\theta-1} \sim L^{-2} 
\end{equation}
where the values $\nu=1/4$ and $\theta=3/2$ were used.
The result $D \sim L^{-2}$ is consistent with the simulations and intuitive arguments of
 Ref \cite {obukhov}. It is remarkable that the diffusion coefficient scaling
 is the same as in a linear reptating chain \cite{de Gennes}, though the microscopic "walker"
 responsible for the motion has very different dynamics.  In the case
of a linear reptating chain, the exponents for the chain size and
walker lifetime, $\nu=1/2$ and $\theta=2$, are different from the ring case,
but they combine to give the same scaling
for $D$. The longest relaxation
 time, in which the ring diffuses about one average ring size, scales as 
$T_{max} \sim L^{5/2}$ (compare to the linear reptating chain result
 $T_{max} \sim L^{3}$) .

The relaxation behavior is very different from that of a linear reptating chain
 \cite {obukhov}. In the linear reptating chain, for most structures to relax, the head of the
 chain must reptate and free them, a process takes of the the order of the longest
 relaxation time $T_{max} \sim L^3$. In contrast, the hernia mechanism allows most
 structures to relax very quickly. 

{\bf 5. Connection with Spontaneous Symmetry Breaking in 1D:} 
In this section, we discuss the models of gel-electrophoresis in the context
 of systems that display spontaneous symmetry breaking (SSB) in one dimension (1D).
 It is well known that in thermal equilibrium, symmetry breaking and long range order
 can not appear in 1D systems with short-range interactions.
 Recently, a simple non-equilibrium model that displays SSB in 1D 
was presented \cite{mukamel}. The Rubinstein-Duke \cite{duke j chem phys}
 reptation-tube model offers an additional interesting example of a 
1D system with short-range interactions which displays symmetry breaking. 
In this model, symmetry breaking has a particularly simple physical meaning.
 Consider a linear DNA chain with no tension or
 hernias. It is typically aligned with the field \cite {duke j chem phys}. There are
 two symmetric configurations: either segment number 1 is 
at the chain head, or segment
 number $L$ is at the chain head. A finite chain eventually flips between these two
 configurations. The flipping time, however, grows very quickly with the chain
 size: in order to flip, the tail must reptate an order 
of $L$ steps against the field,
 and the rest of chain must follow it, and thus the flipping time goes as 
$\sim e^{\alpha L^2}$ with some constant $\alpha$. 
The chain is therefore effectively stuck in one of the two
 configurations and symmetry is broken in the thermodynamic limit.

In this example, as in the model of Ref \cite{mukamel}, the mechanism permitting 
spontaneous symmetry breaking
 depends on the presence of boundaries (i.e. the existence of a chain head and tail)
 and on a conservation law in the chain bulk (the number of $+$ and $-$ particles
is conserved by the bulk dynamics when hernias are not permitted). 
The latter is seen to be important
 by allowing hernias, as in the present model 
(without tension which corresponds to long ranged interactions). Hernias amount to 
pair creation and annihilation in the bulk, and break the conservation law.
 Adding hernias also cancels the SSB: the chain is typically stuck in a 
symmetric hooked state (Sec 3a., Fig 5).

 We note that an interesting case of SSB in a 1D system, 
with periodic boundary conditions and with no bulk conservation of the 
order parameter has been presented \cite{growth}.

\noindent
{\bf 6. Discussion}

The role of tension and hernias in the dynamics of ring-shaped DNA gel-electrophoresis is
 studied. A microscopic model, which adds tension and hernias to the reptation model 
of Rubinstein and Duke, was proposed. The model is similar to reaction-diffusion models,
 with the addition
 of a new and interesting hierarchical pairing which represents the hernia fingers.

 We predict that
 the mobility of open-circular DNA should saturate much as in linear chains. Tension
 serves to stabilize a quasi-linear conformation of the ring. In this regime, there 
are many coiled DNA segments, screening the effects of tension.
 The screening of tension by coiled segments is 
essential for explaining the success of reptation
 models for linear DNA fragments \cite{biopolymers,schutz}, 
which neglect both hernias and tension. It is suggested
that tension and hernias have, to a certain extent, canceling effects.
 It is thus important to include both effects in models of hernia-mediated polymer
 dynamics. Some of the qualitative features of the motion of the ring are captured by a simple
mean field theory.  

The present predictions are difficult to check in standard electrophoresis experiments
 in agarose gels in which the DNA rings become hooked on dangling ends in the gel
 ("hoop in stick"  effect, Sec 3b). An experimental situation which eliminates the effects
 of dangling ends is offered by the microlitthographic arrays of posts introduced
 by Volkmuth and Austin \cite{austin}. These arrays were used to study 
electrophoresis of linear DNA, which could be observed by fluorescence microscopy.
 Since these 2D arrays have a floor and a ceiling attached to the posts,
 a ring-shaped DNA can not become impaled by a post. Thus the hoop in stick effect is negated.
 The present work predicts that the mobility of plasmids in such an obstacle array
 should saturate at a finite value, and not decrease exponentially with the ring
 size as in standard gels with dangling ends.

 In addition,
agarose gels exhibit a very broad distribution of pore sizes.
The present model pertains more closely to experiments on regular lattice, 
since the broad distribution of pore sizes
in standard agarose gels might well affect the dynamical features and
scaling laws discussed in this paper.

The model is also applied to study the dynamics in zero field. The solution of the
 model in this case allows an exact foundation of some known
scaling properties of ring
 polymers. The main effect of hernias on the static chain properties are to allow
 many undulations and wiggles per unit contour length. As a result, the ring adopts
 much more compact configurations (average size $R \sim L^{1/4})$
 than in the reptation case (where $R \sim L^{1/2}$).
 The dynamics have been studied using a new approximate equation of motion for
 marked hernias. This equation seems to be a useful basis for analytical
study of the effect of hernias on polymer dynamics.
The lifetime of a hernia was found to scale as $\tau \sim L^{3/2}$, which
is shorter than expected from pure diffusion $\tau \sim L^2$. 
This is due to an entropic
 bias of small hernias to shrink and be annihilated.
The diffusion coefficient is found to scale as 
in the reptation case, $D \sim L^{-2}$,
though the behavior of the microscopic walker
is different in the reptation and ring cases.
 The relaxation spectrum is very different, with most structures relaxing on 
short time-scales. These results were obtained for ring-shaped polymers, but should
 also be applicable to long linear chains in which hernia creation in the bulk is
 allowed. An exact solution of the steady state is also possible in this case, 
 showing a crossover between reptation and hernia behavior as a function of polymer
 length. This solution, as well as a more detailed study of the zero-field dynamics,
 will be presented in a future publication.

Finally, we note that the physically intuitive treatment of chain tension presented here
may be useful also for describing different systems in which polymers in a melt or 
obstacle array move under external forces.

Acknowledgment: We thank R. Austin, S. Leibler, S. Sandow and 
L. Ulanovski for helpful 
discussions. This work was supported by the Minerva Foundation, Munich.

%\vspace{-5mm}

%-------------------------------- figure 1 ----------------------------------

\begin{figure}
      
   \caption { Configuration of a DNA chain (heavy line) in a gel, defined by a periodic lattice of gel
 pores (dotted lines). The DNA is divided into persistence length segments, numbered $i=1,2 \cdots L$ (in this 
case $L=8$). The reptation tube (light line) is defined by all pores through which the DNA threads. The configuration is
 encoded using a $+$ for segments that are stretched between two pores with the field direction, $-$ for
 segments stretched against the field direction, and $0$ for coiled segments in the same pore. The displayed
 configuration is thus $-0++00-+$ for $i=1 \cdots 8$. Move A corresponds to to a standard reptation move
 (inside the reptation tube) $+0 \rightarrow 0+$. This move occurs with rate $p$. The reverse move
 $0+ \rightarrow +0$ is against the field direction and is given a smaller rate $q$. Move B corresponds
 to the formation of a hernia (leak through the reptation tube). It is represented by pair creation
 $00 \rightarrow +_{\sqcup}-$. } 
\label {fig dna gel}
\end{figure}

%------------------------------------figure 2----------------------------------

\begin{figure}
 \caption { A configuration of a ring-shaped DNA in the gel. The segments are numbered counter-clockwise.
 The 
configuration contains several branching fingers, and is encoded as shown by a string of $+$,$-$ and
 $0$'s along with their pairings into hernias . Ring-shaped DNA can move through the gel only by
 the annihilation and creation of hernias.} \label {fig dna ring}
\end{figure}
\noindent

%--------------------------------- figure 3
\begin{figure}
 \caption {An example of a configuration of $+$,$-$ and $0$ charges in the model. a) One of
the allowed pairings into hernias. b) A forbidden pairing, where the hernias are
not nested. Such a configuration can not be reached from an allowed configuration
by the model dynamics.} 
\end{figure}
\noindent

%------------------------------------figure 4----------------------------------

\begin{figure}
\caption {Monte-Carlo simulation results of the model including hernias but neglecting tension,
for a ring of size $L=52$, with $p=0.56$,
$q=0.5$ and $c=a=1$.
The mobility of the ring (center of mass velocity)
 is shown as a function of time (in sweeps, where
 one sweep equals $L$ single-bond moves). The mobility is seen to have a spiked behavior, where the
 mobility is mostly zero with intermittent periods of motion.}
 \label{fig mc}
\end{figure}
\noindent

%------------------------------------figure 5----------------------------------

\begin{figure}
 \caption {The ring configurations between one mobile burst and the next. The ring begins with a quasi-linear shape aligned with the field,
 which quickly develops an instability to secondary hernias and goes to a 2-armed hook. The hooks (in the 
absence of tension) are stuck for long times. Unhooking occurs by the retraction of one of the arms by 
fluctuations against the field, until a quasi-linear shape with a non-zero mobility is reached again.
 The spikes in the ring mobility in Fig. 3 are thus explained by the cycle between quasi-linear and
 hooked conformations.}
 \label{fig cycle}
\end{figure}
\noindent
%------------------------------------figure 6----------------------------------
\begin{figure}
 \caption 
{Configuration of charges and hernia-pairings in the model that corresponds to a) Quasi-linear conformation, 
b) Hook with two equal arms. The quasi-linear conformation typically also includes many 0 sites, 
as well as small branching sub-hernias.}
 \label{fig hernia}
\end{figure}
\noindent

%------------------------------------figure 7----------------------------------
\begin{figure}
        
 \caption {
Mobility of a ring-shaped DNA fragment as a function of size. Monte-Carlo results for two field strengths, $E=2$ (light lines) and $E=1$ (bold lines) are shown with hernia creation and annihilation rates
 $a=e^{-1}$ and $c=0.3$. Units are such that $Qb/kT=1$ so that the dimensionless field $\epsilon=E$.
 The mobility of the model including tension (full lines) decreases with 
the chain length for short chains, and then shows a saturation. For stronger fields, the asymptotic
 mobility is higher, and the chain length $L ^{\star}$ at which the mobility saturates is smaller.
 The results of the model without tension (dashed lines) are close to the results with tension for
 short chains ($L <L^{\star}$). However, without tension, an exponentially decreasing mobility is
 predicted for long chains, because of the formation of hooks.} \label{fig mu}
\end{figure}
\noindent
%------------------------------------figure 8----------------------------------

\begin{figure}
 \caption {
 Density of coiled segments, $\rho$, and mobility, $\mu$, of ring-shaped DNA as a function of the 
ratio of hernia creation and annihilation rates $c/a$. The ring size is $L=40$, 
the field strength is $E=2$ and the annihilation rate is held constant $a=e^{-1}$. 
Shown are Monte-Carlo simulation results of the model including tension (o's), 
and the mean-field prediction (line). } 
\label{fig cr}
\end{figure}
\noindent


\begin{thebibliography}{10}
\bibitem{open circular} The term open circular refers to ring-shaped DNA which
is not supercoiled (closed circular). Open-circular DNA is obtained from supercoiled
DNA by a nicking procedure, which removes base-pairs from one of the strands,
allowing the double-stranded DNA to swivel and release the torsion which caused it to 
supercoil \cite{mickel}.
\bibitem{quasi} The quasi-linear conformation is represented in the model
by a phase-separated distribution of charges, where half of the lattice is
rich with $+$ sites and the other half rich with $-$ sites,
as shown in Fig. 6a, along with the appropriate pairing into hernia pairs. Typically 
the conformation also includes many $0$ sites, small branching sub hernias etc.
\bibitem{sign} This equation represents the positive root of a quadratic 
equation for $g(s)$. The negative root corresponds to a non-physical solution
in which $N(L)$ is negative at large $L$.
\end{thebibliography}
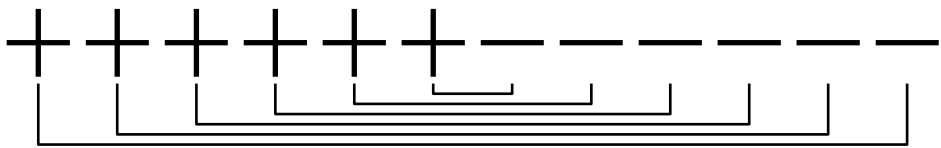
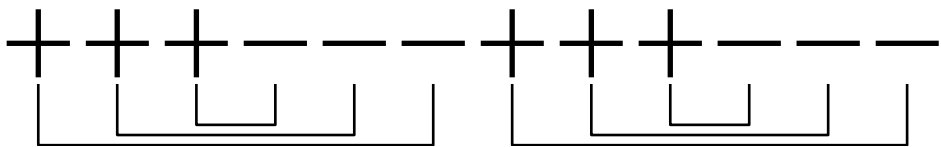
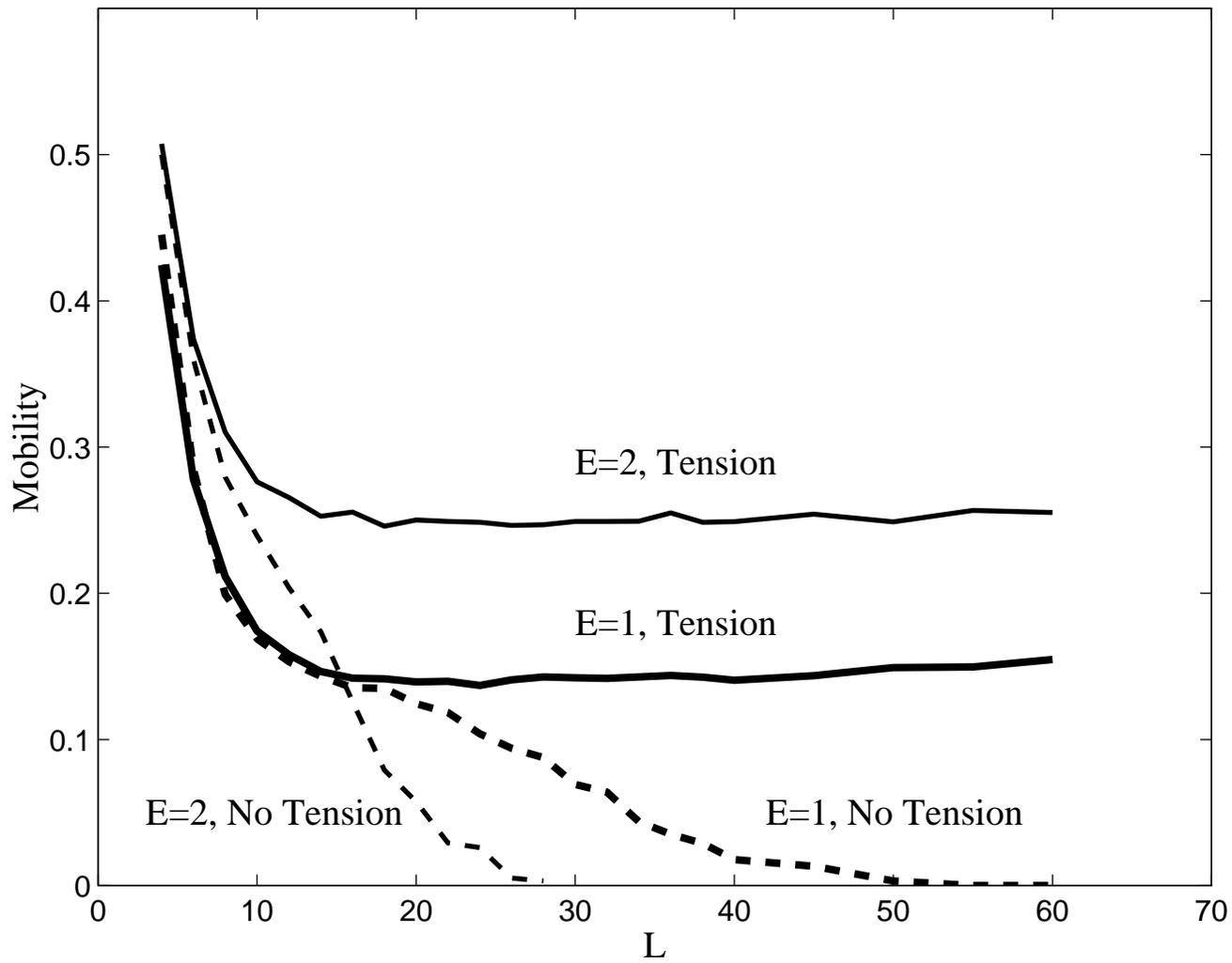
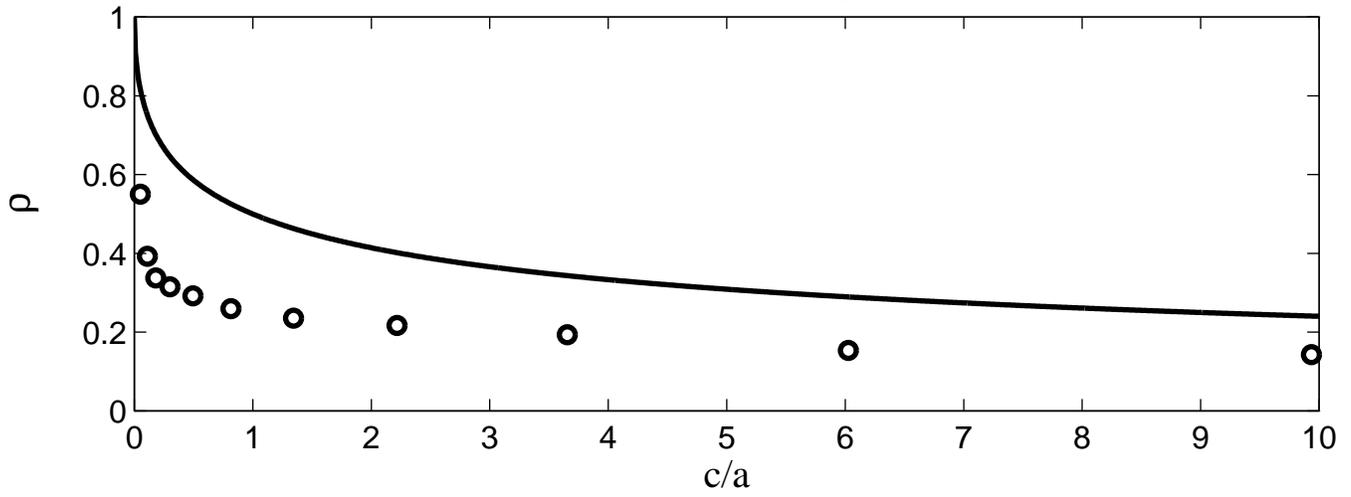
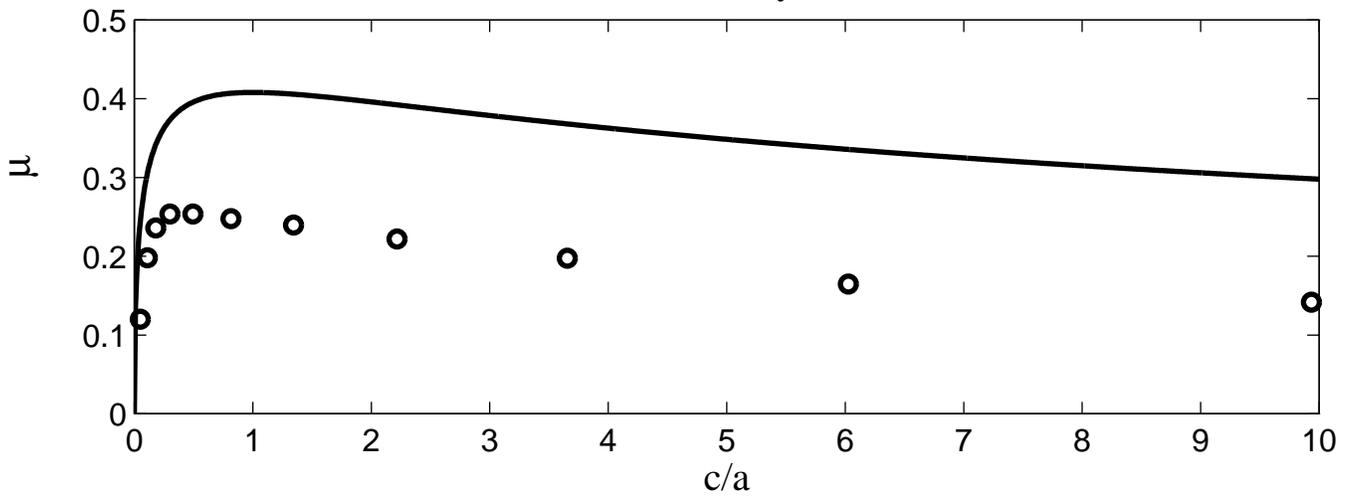
\end{document}